\documentclass[11pt]{article}
\usepackage{graphicx}
\usepackage{amssymb}

\begin{document}

\title{ \bf \huge Boundary Terms in Supergravity and Supersymmetry  \footnotemark}

\author{Peter van Nieuwenhuizen \\ [5pt]
{\it C.N. Yang Institute for Theoretical Physics} \\
{\it Stony Brook University, Stony Brook, NY 11794-3840, USA} \\
\texttt{vannieu@insti.physics.sunysb.edu} \\ [5pt]
Anton Rebhan  \\ [5pt]
{ \it Institut f\"ur Theoretische Physik, Technische Universit\"at Wien,} \\
{\it Wiedner Hauptstr. 8--10, A-1040 Vienna, Austria} \\ [5pt]
Dmitri V.Vassilevich \\ [5pt]
{\it Institut f\"ur Theoretische Physik, Universit\"{a}t Leipzig} \\
{\it Augustusplatz 10, D-04109 Leipzig, Germany} \\ [5pt]
Robert Wimmer \\ [5pt]
{\it Institut f\"ur Theoretische Physik, Universit\"at Hannover,} \\
{\it Appelstr.~2, D-30167 Hanover, Germany}\\ [10pt]} 

\bigskip


\maketitle

\renewcommand{\thefootnote}{\fnsymbol{footnote}}

\footnotetext[1]{Talk given by P. van Nieuwenhuizen at the Einstein-celebration gravitational conference at Puri (India) in December 2005.}

\renewcommand{\thefootnote}{\arabic{footnote}}

{\bf Abstract:}{\ We begin with the simplest possible introduction to supergravity. Then we discuss its spin 3/2 stress tensor; these results are new. Next, we discuss boundary conditions on fields and boundary actions for $N=1$ supergravity. Finally, we discuss new boundary contributions to the mass and central charge of monopoles in $N=4$ super Yang-Mills theory. All models are in $3+1$ dimensions.
\begin{picture}(0,0)
\put(306,450){YITP-SB-06-21}
\end{picture}
}

\section{Introduction}

Supergravity is 30 years old. It is an active field as several of the speakers at this conference demonstrate. I will present a new program in supergravity, with some initial results, but much more can be done. For those who are not familiar with supergravity, I will give the simplest introduction to supergravity I know of; it is not the way supergravity was originally constructed \cite{sugra}, but it is how it was simplified years later, bit by bit. It is suitable to become part of a course on general relativity.

Boundary contributions to the variation of the action are usually neglected in theories of rigid or local supersymmetry (susy). However, a theory is only complete if one has also defined its boundary conditions. In this contribution we shall discuss this problem; one may think of putting supergravity in a box. Of course, boundary conditions play a central role in such classical solutions as instantons and solitons. As a second topic, we study how boundary terms also play a role in the study of solitons at the quantum level: we shall discuss the quantum contributions to the mass of magnetic monopoles in $N=2$ and $N=4$ super Yang-Mills theory. 

The main new point in our discussion of boundary conditions (BC) on fields concerns their {\it consistency}. Given an action with a (rigid or local) symmetry, there are boundary terms in the variation of the action (for spacetime symmetries, including supersymmetry, but not for internal symmetries), which should cancel if the symmetry is unbroken. This one can achieve by imposing suitable BC, or by adding a boundary term to the action (boundary action), or by both. Another class of boundary variations is obtained when one constructs the Euler-Lagrange field equations. Also in this case one needs to partially integrate, and the boundary terms obtained in this way are in general different from those obtained from the variation of the action under a symmetry transformation. By definition, these boundary field equations should be satisfied on-shell. We shall actually require that they be satisfied also off-shell because we want to characterize the field space by a complete set of boundary conditions, before one begins by studying symmetries of the action and field equations.

However, these are not all BC. If a theory has a symmetry, the symmetry transformation of a BC should also be satisfied, because a theory with fields $\phi$ is invariant under a symmetry transformation $\phi  \longrightarrow \tilde{\phi}$ if nothing changes when written in terms of $\tilde{\phi}$. So the BC on $\phi$ must be the same as on $\tilde{\phi}$, and this implies that a symmetry transformation of a given BC must again be a BC. If this new BC is not part of the set of BC obtained at this point, we have to add it. In this way we obtain a whole 'orbit' of BC. So the BC we need come from 
\begin{itemize}
\item[(i)] Euler-Lagrange field equations 
\item[(ii)] symmetries of the action 
\item[(iii)] variations thereof 
\item[(iv)] conjugate BC
\end{itemize}
The last set, conjugate BC, occur when some bulk field equations vanish at the boundary as a consequence of the BC. One must then impose these field equations which are 'conjugate' to the boundary conditions separately as BC. There are many articles in the literature  of general relativity on BC \cite{Luckock:1989jr,D'Eath:1984sh,D'Eath:1996at,Esposito:1996kv,Luckock:1990xr,Branson:1999jz,Eath:1991sz,EKPbook,HennTeit}, but the concept of their consistency is new, as far as I know, and was developed by Ro\v{c}ek, Lindstr\"{o}m and myself \cite{Lindstrom:2002mc}. This approach was tested at the example of susy kink in \cite{Bordag:2002dg}. One can also use an approach based on rigid superspace methods \cite{Belyaev:2006} to determine a boundary action in $x$ space such that the rigid susy of the total action does not lead to BC on the fields (while the boundary field equations still lead to BC on the fields which form a closed orbit under 
rigid susy). For a discussion of susy junction conditions see \cite{Moss:2005}.

In ordinary (Einstein-Hilbert) gravity, boundary terms are well-known from the work of Gibbons, Hawking and York \cite{Gibbons:1976ue,York:1972sj}. These authors noted (it is only a brief remark in the Gibbons-Hawking paper) that if one adds the extrinsic curvature as boundary action on a boundary $\partial {\cal M}$ of spacetime ${\cal M}$, then under an Euler-Lagrange variation one finds the following result
\begin{equation}
\delta S_{EH} + \delta S_{bound} = \int_{\cal M} G^{\mu\nu} \delta g_{\mu\nu} + \int_{\partial {\cal M}} K^{ij} \delta g_{ij}
\end{equation}
So the bulk field equations state that the Einstein tensor $G^{\mu\nu}$ must vanish, but the boundary field equations require $K^{ij} \delta g_{ij}$ to vanish, where $g_{ij}$ is the metric in the surface, and $K^{ij}$ the extrinsic curvature. Gibbons and Hawking proposed to impose as BC that $\delta g_{ij} = 0$. With Vassilevich \cite{Nieuwenhuizen:Vassilevich:2005} I have studied the case of supergravity, and found that the condition $\delta g_{ij} = 0$ violates local supersymmetry. Instead we found a consistent orbit of BC which contains the BC  $K^{ij}=0$.

The program of studying BC in general field theories while taking consistency into account is a large program, and only a few initial results have been obtained. For example, in superspace one often replaces Grassmann integration by differentiation with susy covariant derivatives $D_{\alpha} = \frac{\partial}{\partial \theta^{\alpha}} + \bar{\theta}^{\dot{\alpha}} \sigma^{\mu}_{\alpha\dot{\alpha}} \partial_{\mu}$. The terms with $\partial_{\mu}$ in $D_{\alpha}$ are supposed to cancel, but clearly this is not true in general when there are boundaries. In fact, one finds that a 'boundary superspace' can be defined \cite{Lindstrom:2002mc,Belyaev:2006}. Of course, BC play an important role in string theory; that was, in fact, where the consistency of BC was first studied \cite{Lindstrom:2002mc}. Also in the AdS-CFT program of string theory, BC as one approaches the boundary play a crucial role \cite{Breitenlohner:Freedman:Hawking}. However, as far as I know, their consistency with local susy has not been studied, and this might be an interesting subject to work out.

As a a second topic we shall discuss the contribution of boundary terms to the mass and central charges of monopoles. For several years three of us (A.R., R.W. and P.v.N.) have studied solitons at the quantum level. We began with kinks, where a new anomaly (in the {\it conformal} currents) was found to contribute to the (non-conformal) central charge (= magnetic charge) of a monopole. Only with this anomaly did BPS saturation hold. Then we turned to vortices; since they are in 2 space dimensions, there cannot be anomalies, but we found a quantum effect in the winding which previous authors had missed, and again, only with this quantum winding did BPS saturation hold. Finally, we turned to monopoles. For the $N=2$ case, the same anomaly as for the kink was needed for BPS saturation. However, in the $N=4$ case, two new effects were discovered: {\it surface terms in the mass}, and {\it composite operator renormalization} of the currents. Looking back at our previous models we noted that these new effects were absent because either these models had no massless fields (kink and vortex), or there occurred 'miraculous' cancellations ($N=2$ monopoles).

\section{Simple supergravity made simpler}

Consider simple ($N=1$) supergravity. The action is the sum of the Hilbert-Einstein action for pure gravity (but in the form of Weyl with  vielbeins $e_{\mu}{}^m$ and a spin connection instead of the normal vector connection) and the Rarita-Schwinger action for the real massless spin $3/2$ gravitino, ${\cal L} = {\cal L}_{EH} + {\cal L}_{RS}$, where
\begin{eqnarray}
{\cal L}_{EH} &\hspace{-5pt} = \hspace{-5pt}& - \frac{1}{2\kappa^2}\, e\, e_m{}^{\nu} e_n{}^{\mu} R_{\mu\nu}{}^{mn}(\omega) \label{L_EH} \\
{\cal L}_{RS} &\hspace{-5pt} = \hspace{-5pt}& -\frac{1}{2}\, e \, \bar{\psi}_{\mu} \gamma^{\mu\rho\sigma} D_{\rho}(\omega) \psi_{\sigma} 
\end{eqnarray}
with
\begin{eqnarray}
R_{\mu\nu}{}^{mn}(\omega) &\hspace{-5pt} = \hspace{-5pt}& \partial_{\mu} \omega_{\nu}{}^{mn} + \omega_{\mu}{}^m{}_k \, \omega_{\nu}{}^{kn} - \mu \leftrightarrow \nu \\
D_{\rho} \psi_{\sigma} &\hspace{-5pt} = \hspace{-5pt}& \partial_{\rho} \psi_{\sigma} +\frac{1}{4} \omega_{\rho}{}^{mn} \gamma_{mn} \psi_{\sigma}
\end{eqnarray}
We used $e = \det e_{\mu}{}^{m}$,\ $\gamma_{mn} = \frac{1}{2}(\gamma_m\gamma_n-\gamma_n\gamma_m)$ with strength one, and $\gamma^{\mu\rho\sigma} = e_{m}{}^{\mu} e_{r}{}^{\rho}e_{s}{}^{\sigma} \gamma^{mrs}$ with $\gamma^{mrs}$ antisymmetric in $m$, $r$, $s$, also with strength one. The Dirac matrices $\gamma^m$ are constant. {\it For the time being we leave $\omega_{\mu}{}^{mn}$ unspecified.}

This action is invariant under the following local susy transformations
\begin{eqnarray}
\delta_{\rm susy} e_{\mu}{}^m &\hspace{-5pt} = \hspace{-5pt}& \frac{\kappa}{2} \bar{\epsilon}\, \gamma^m \psi_{\mu} \label{var_vielbein} \\
\delta_{\rm susy} \psi_{\mu} &\hspace{-5pt} = \hspace{-5pt}& \frac{1}{\kappa} D_{\mu}(\omega) \epsilon \label{var_gravitino} \\ [7pt]
\delta_{\rm susy} \omega_{\mu}{}^{mn} &\hspace{-5pt} = \hspace{-5pt}& ?
\end{eqnarray}
where $\delta_{\rm susy} \omega_{\mu}{}^{mn}$ is to be determined once $\omega_{\mu}{}^{mn}$ has been specified. The proof of invariance proceeds in two steps. 

{\bf Step 1}: the variation of $\psi_{\sigma}$ and $\bar{\psi}_{\mu}$ in ${\cal L}_{RS}$ cancels the variation of $e\, e_m{}^{\mu} e_n{}^{\nu}$ in ${\cal L}_{EH}$. Any variation of the vielbeins in ${\cal L}_{EH}$ is proportional to the Einstein tensor $R_{\mu}{}^n - \frac{1}{2} e_{\mu}{}^n R$ where $R_{\mu}{}^n \equiv R_{\mu\nu}{}^{mn}(\omega) e_m{}^{\nu}$. On the other hand, the commutators $[D_{\rho}, D_{\sigma}] \epsilon$  and (after partial integration) $[D_{\mu}, D_{\rho}] \psi_{\sigma}$ in the variation of ${\cal L}_{RS}$ yield two terms with curvatures which combine into an anticommutator
\begin{equation}
-\frac{e}{16 \kappa} R_{\rho\sigma}{}^{mn} \bar{\psi}_{\mu} \{ \gamma^{\mu\rho\sigma}, \gamma_{mn} \} \epsilon \label{anticomm}
\end{equation}
To arrive at this expression one needs to use the identity $\bar{\epsilon} \gamma_{m_1} ... \gamma_{m_n} \psi_{\sigma} = (-)^n \bar{\psi}_{\sigma} \gamma_{m_n} ... \gamma_{m_1} \epsilon$, which is valid for Majorana spinors $\psi_{\sigma}$ and $\epsilon$. Each of the products $\gamma^{\mu\rho\sigma} \gamma_{mn}$ and $\gamma_{mn} \gamma^{\mu\rho\sigma}$  is a sum of totally antisymmetrized terms with 5, 3 and 1 gamma matrices, but in the anticommutator in (\ref{anticomm}) the 3-gamma terms cancel, and the 5-gamma and 1-gamma terms add up. Moreover, the terms with 5 gamma matrices totally antisymmetrized in their vector indices vanish in four dimensions (because there are always at least two indices equal). One is then left in (\ref{anticomm}) with terms of the form $\bar{\epsilon}  \gamma \psi$ times again the Einstein tensor. Requiring that the sum of the two coefficients of the Einstein tensor vanishes, {\it one derives the variation of $e_{\mu}{}^m$} in (\ref{var_vielbein}) in this way. Crucial for our purposes is the boundary term one obtains by partially integrating $\delta \bar{\psi}_{\mu} = \frac{1}{\kappa} D_{\mu} \bar{\epsilon}$; it clearly reads
\begin{equation}
\partial_{\mu} \left[- \frac{e}{2\kappa} \bar{\epsilon} \gamma^{\mu\rho\sigma} D_{\rho}(\omega) \psi_{\sigma} \right] \label{bound_term}
\end{equation}

{\bf Step 2}: One is left with the following four variations
\begin{itemize}
\item[(i)] the variation of $\omega$ in ${\cal L}_{EH}$ 
\item[(ii)] the variation of $\omega$ in ${\cal L}_{RS}$
\item[(iii)] the variation of the vielbein in ${\cal L}_{RS}$
\item[(iv)] the terms with $D_{\mu} e_{\nu}{}^m$ which are obtained if one partially integrates $D_{\mu}\bar{\epsilon}$
\end{itemize}
To simplify the evaluation of (iii) and (iv), we use two other ways to write the action
\begin{eqnarray}
{\cal L}_{EH} &\hspace{-5pt} = \hspace{-5pt}& -\frac{1}{8\kappa^2} \epsilon^{\mu\nu\rho\sigma} \epsilon_{mnrs} R_{\mu\nu}{}^{mn}(\omega) e_{\rho}{}^r e_{\sigma}{}^s \label{L_EH_alternative} \\
{\cal L}_{RS} &\hspace{-5pt} = \hspace{-5pt}& -\frac{i}{2} \epsilon^{\mu\nu\rho\sigma}\bar{\psi}_{\mu} \gamma_5 \gamma_{\nu} D_{\rho} \psi_{\sigma} \label{L_RS_alternative}
\end{eqnarray}
In (\ref{L_EH_alternative}), the product of the two $\epsilon$-tensors gives a product of vielbeins, namely $\epsilon^{\mu\nu\rho\sigma} \epsilon_{mnrs} e_{\rho}{}^r e_{\sigma}{}^s = -2\, e\, (e_m{}^{\mu} e_n{}^{\nu} - e_n{}^{\mu} e_m{}^{\nu})$, and yields back (\ref{L_EH}). Further, we used in (\ref{L_RS_alternative}) that the product $\gamma^{mrs}$ is equal to $i \epsilon^{mrst} \gamma_5\gamma_t$. \footnote{We define $\gamma_5 = \gamma^1 \gamma^2 \gamma^3 i \gamma^0$ with $(\gamma_5)^2 = (\gamma^k)^2=I$ but $(\gamma^0)^2 = -I$. Note that $(\gamma^k)^{\dagger} = \gamma^k$ and $(\gamma_5)^{\dagger} = \gamma_5$ but $(\gamma^0)^{\dagger} = -\gamma^0$. Also, $\bar{\psi}_{\mu} = \psi_{\mu}^{\dagger} i \gamma^0 = \psi_{\mu}^T C$ where $C\gamma_m C^{-1} = - \gamma_m^T$. We use the convention $\epsilon^{0123} =1=-\epsilon_{0123}$.}
The technical advantage of (\ref{L_RS_alternative}) is that it contains only one vielbein field (since $\epsilon^{\mu\nu\rho\sigma}$ is a density, there is no factor $e$). It is straightforward to evaluate these four variations, and they factorize (!)
\begin{equation}
\delta (\mbox{remaining}) {\cal L} = \epsilon^{\mu\nu\rho\sigma} \epsilon_{mnrs} \frac{1}{2\kappa^2} (\delta_{\rm susy} \omega_{\mu}{}^{mn} e_{\nu}{}^r + \frac{\kappa}{6} \bar{\epsilon} \gamma^{mnr} D_{\mu} \psi_{\nu} ) (D_{\rho} e_{\sigma}{}^s - \frac{\kappa^2}{4} \bar{\psi}_{\rho} \gamma^s \psi_{\sigma}) \label{remaining_var}
\end{equation} 
One now easily understands the two versions of supergravity: 

(A) The second-order formalism according to which
\begin{equation}
D_{[\rho} e_{\sigma]}{}^s = \frac{\kappa^2}{4} \bar{\psi}_{[\rho}\gamma^s \psi_{\sigma]} \label{De}
\end{equation}
One can solve this equation for $\omega_{\mu}{}^{mn}$ and finds then
\begin{equation}
\omega_{\mu}{}^{mn} = \omega_{\mu}{}^{mn} (e) + \frac{\kappa^2}{4} (\bar{\psi}_{\mu} \gamma^m \psi^n - \bar{\psi}_{\mu} \gamma^n \psi^m + \bar{\psi}^m \gamma_{\mu} \psi^n) \label{supercov_spin_conn}
\end{equation}
where $\omega_{\mu}{}^{mn} (e)$ is the usual textbook spin connection, a composite field depending on $e_{\mu}^m$ 
\begin{equation}
\omega_{\mu}{}_{mn} (e) = [ \frac{1}{2} e_m{}^{\nu} (\partial_{\mu} e_{n\nu} - \partial_{\nu} e_{n\mu}) - m \leftrightarrow n ] - \frac{1}{2} e_m{}^{\rho} e_n{}^{\sigma} (\partial_{\rho} e_{\sigma}{}^c - \partial_{\sigma} e_{\rho}{}^c) e_{c\mu} \label{spin_conn}
\end{equation}
and the $\bar{\psi} \gamma \psi$ terms are torsion. This is the solution of Freedman et al. in \cite{sugra}. 

(B) The first-order formalism in which $\omega_{\mu}{}^{mn}$ is an independent field, whose variation is given by requiring the first factor in (\ref{remaining_var}) to vanish. One can solve for $\delta_{\rm susy} \omega_{\mu}{}^{mn}$ the same way as one solves for $\omega_{\mu}{}^{mn}$ and finds then
\begin{equation}
\delta_{\rm susy} \omega_{\mu}{}_{mn} = -\frac{1}{2} \bar{\epsilon}\, \gamma_5 \gamma_{\mu} \tilde{\psi}_{mn} + \frac{1}{4} \bar{\epsilon}\, \gamma_5 (\gamma^{\lambda} \tilde{\psi}_{\lambda n} e_{m\mu} - m \leftrightarrow n) \label{var_spin_conn}
\end{equation}
where $\tilde{\psi}_{mn} = \frac{1}{2} \epsilon_{mn}{}^{rs} \psi_{rs}$ and $\psi_{\mu\nu} = D_{\mu}(\omega) \psi_{\nu} - D_{\nu}(\omega) \psi_{\mu}$. This is the solution of Deser and Zumino in \cite{sugra}. 

The factorization in (\ref{remaining_var}) shows that there are only two formulations of supergravity in $x$-space. In superspace one uses second-order formalism, but then one needs to impose by hand constraints on some of the supertorsions (or supercurvatures) which are not field equations. One of the constraints on the components of the supertorsion tensor in superspace is $D_{[\rho} e_{\sigma]}{}^s - \frac{\kappa^2}{4} \bar{\psi}_{[\rho} \gamma^s \psi_{\sigma]} = 0$\,; thus vanishing supertorsion in superspace implies non-vanishing torsion in $x$-space. The expression for $\omega_{\mu}{}^{mn}$ in terms of $e_{\mu}{}^m$ and $\psi_{\mu}$ as given in (\ref{supercov_spin_conn}) is the Euler-Lagrange field equation. {\it Thus we may neglect any variation of $\omega$ in the bulk action} (but not on the boundary, see below). The expression in (\ref{supercov_spin_conn}) is {\it supercovariant}: its susy variation contains no terms with $\partial_{\mu} \epsilon$.  In 10 and 11 dimensions one can repeat this analysis, but there one finds that the supercovariant spin connection differs from the solution of its field equation by terms involving $\bar{\psi}_{\mu} \gamma^{\mu\alpha\beta\gamma\nu} \psi_{\nu}$. 

By far the simplest way to deal with supergravity models is to combine the virtues of (A) and (B) into what has been called the 1.5 order formalism: use $\omega = \omega(e,\psi)$  in (\ref{supercov_spin_conn}) but never expand $\omega(e,\psi)$. Variations of $\omega_{\mu}{}^{mn}(e,\psi)$ in the action are complicated, but they are always multiplied by the $\omega$ field equation, which vanishes. If one solves for $\omega_{\mu}{}^{mn}$, and inserts the resulting $\omega_{\mu}{}^{mn}(e,\psi)$ into the first-order transformation laws, one finds that the difference of the first-order and second-order transformation laws is a separate local symmetry, proportional to the $\omega_{\mu}{}^{mn}$ and $\psi_{\mu}$ field equations (a so-called equation of motion symmetry). 

The 1.5 order formalism is nothing else than the Palatini trick of general relativity extended to supergravity. The transformation laws in (\ref{var_vielbein}) and (\ref{var_gravitino}) can also be obtained by 'gauging' the super Poincare algebra. The first who gauged it were Volkov and Soroka \cite{Volkov:Soroka} who used the formalism of E. Cartan of one-forms, and proposed an action for supergravity. They were studying the Higgs effect for Goldstone particles with spin 1/2. However, because in the super Poincare algebra one has $\{Q,Q\} \sim P_{\mu}$ and not $\{Q,Q\} \sim P_{\mu}+M_{\mu\nu}$ (where $M_{\mu\nu}$ are the Lorentz generators), they found $\delta_{susy} \omega_{\mu}{}^{mn} = 0$, and not (\ref{var_spin_conn}). As a result there is no upper bound $N\leq8$ on the number of possible supergravities in their approach. Later Volkov explained his idea more fully in a CERN publication \cite{Volkov}. The first clear statement about the 1.5 order formalism in supergravity appeared in articles by Chamseddine and West and Townsend and van Nieuwenhuizen \cite{Chamseddine:West,Townsend:Nieuwenhuizen}. The factorization in (\ref{remaining_var}) was found by P.K. Townsend in his 1982 Kyoto lectures (unpublished).

\section{The spin 3/2 stress tensor}

One issue that might be confusing is how to calculate the gravitational stress tensor for spin 3/2 fields using the 1.5 order formalism. The 1.5 order formalism shows that one need not vary the vielbeins in the spin connection of {\it both} ${\cal L}_{EH}$ and ${\cal L}_{RS}$ because the sum of all these variations anyhow cancels. Thus the spin 2 field equation is simply obtained by varying only the explicit vielbeins in the action, and reads
\begin{equation}
\frac{e}{\kappa^2} (R_{\nu}{}^{\tau} - \frac{1}{2} \delta_{\nu}{}^{\tau} R) = \frac{i}{2} \bar{\psi}_{\mu} \gamma_5 \gamma_{\nu} D_{\rho} \psi_{\sigma} \epsilon^{\mu\tau\rho\sigma} \equiv \theta_{\nu}{}^{\tau} \label{spin2_eom}
\end{equation}
The spin 3/2 field equation can be obtained, using again the 1.5 order formalism, by only varying the explicit gravitino fields in ${\cal L}_{EH}+{\cal L}_{RS}$, and reads  
\begin{equation}
\epsilon^{\mu\nu\rho\sigma} (\gamma_5 \gamma_{\nu} D_{\rho}  \psi_{\sigma} + \frac{1}{2} \gamma_5 \gamma_n \psi_{\sigma} D_{\rho} e_{\nu}{}^n) = 0
\end{equation}
The last term actually vanishes since $D_{[\rho} e_{\nu]}{}^n = \frac{\kappa^2}{4} \bar{\psi}_{[\rho} \gamma^n \psi_{\nu]}$ and $(\gamma_n \psi_{\sigma}) (\bar{\psi}_{\rho} \gamma^n \psi_{\nu}) \epsilon^{\mu\nu\rho\sigma} = 0$. So, the complete spin 3/2 field equation in 1.5 order formalism is simply
\begin{equation}
R^{\mu} \equiv \epsilon^{\mu\nu\rho\sigma} \gamma_5 \gamma_{\nu} D_{\rho}(\omega) \psi_{\sigma} = 0 \label{gravitino_eom}
\end{equation}
where $\omega = \omega(e,\psi)$. The consistency condition $D_{\mu}R^{\mu} = 0$ is satisfied (see the second paper in \cite{sugra}). To prove this it is easiest to use (\ref{L_RS_alternative}), (\ref{De}) and (\ref{spin2_eom}) and the torsion equation
\begin{equation}
\epsilon^{\mu\nu\rho\sigma} R_{\mu\nu\rho s} = \kappa^2 \bar{\psi}_{\mu} \gamma_s D_{\nu} \psi_{\rho} \epsilon^{\mu\nu\rho\sigma}
\end{equation}
(always with $\omega = \omega(e,\psi)$). Tracing the spin 2 field equation produces the spin 3/2 field equation on the right-hand side, so on-shell the scalar curvature vanishes
\begin{equation}
R = 0
\end{equation}

To obtain the classical gravitational spin 3/2 stress tensor, one may begin by writing the action as follows
\begin{equation}
{\cal L}_{EH} (e,\omega(e,\psi)) + {\cal L}_{RS} (e,\psi, \omega(e,\psi)) = {\cal L}_{EH} (e,\omega(e)) + {\cal L}_{RS} (e,\psi, \omega(e)) + {\cal L}_{to} \label{action}
\end{equation}
where the torsion terms are given by (see the review in \cite{sugra})
\begin{equation}
{\cal L}_{to} = \frac{e}{2\kappa^2} \left[ \tau_{\mu\nu\rho} \tau^{\rho\nu\mu} - (\tau^{\lambda}{}_{\lambda\mu})^2 \right]
\end{equation}
with $\tau_{\mu\nu\rho} = \omega_{\mu\nu\rho}(e,\psi) - \omega_{\mu\nu\rho}(e)$ the torsion terms in (\ref{supercov_spin_conn}). The variation of the last two terms in (\ref{action}) with respect to the vielbein field gives then the spin 3/2 stress tensor. It contains, in addition to $\theta_{\nu}{}^{\tau}$, total derivative terms with two gravitinos due to varying $\omega(e)$, and four-gravitino terms from ${\cal L}_{to}$. 

Note that $\theta_{\nu}{}^{\tau}$ is not proportional to the stress tensor $T_{\nu}{}^{\tau}$ because the terms due to varying the vielbeins in the spin connection of ${\cal L}_{RS}$ and in ${\cal L}_{to}$ are missing. The stress tensor of an Einstein- and locally Lorentz-invariant matter action should be symmetric and covariantly conserved on the spin 3/2 mass shell, as follows from
\begin{equation}
\delta S = \int \delta e_{\mu}{}^m T_m{}^{\mu} d^4x = \int D_{\mu} \zeta^m\, T_m{}^{\mu} d^4x = - \int \zeta^m D_{\mu} T_m{}^{\mu} d^4x = 0
\end{equation}
This implies
\begin{equation}\label{eq:26}
D_{\mu}(\omega(e)) T_m{}^{\mu} = 0
\end{equation}
where $\zeta^m = \zeta^{\mu} e_{\mu}{}^m$. (The variation $\delta e_{\mu}{}^m = D_{\mu}(\omega(e)) \zeta^m$ is a sum of an Einstein transformation $\delta e_{\mu}{}^m = (\partial_{\mu} \zeta^{\nu} e_{\nu}{}^m) + \zeta^{\nu} \partial_{\nu} e_{\mu}{}^m$ and a local Lorentz transformation $\delta e_{\mu}{}^m = (\zeta^{\nu} \omega_{\nu}{}^m{}_n) e_{\mu}{}^n$, namely 
\begin{eqnarray}
\delta e_{\mu}{}^m &=& \partial_{\mu} \zeta^m - \zeta^{\nu} ( \partial_{\mu} e_{\nu}{}^m - \partial_{\nu} e_{\mu}{}^m) + \zeta^{\nu} \omega_{\nu}{}^m{}_n e_{\mu}{}^n \nonumber\\
&=& D_{\mu}(\omega(e)) \zeta^m + \zeta^{\nu} (-\partial_{\mu} e_{\nu}{}^m + \partial_{\nu} e_{\mu}{}^m + \omega_{\nu}{}^m{}_n e_{\mu}{}^n -  \omega_{\mu}{}^m{}_n e_{\nu}{}^n) = D_{\mu} \zeta^m
\end{eqnarray}
because $D_{\mu} e_{\nu}{}^m - D_{\nu} e_{\mu}{}^m = 0$ if we take $\omega=\omega(e)$ in (\ref{spin_conn})). Applying this to the spin 3/2 field shows that $D_{\mu}(\omega(e)) T_m{}^{\mu} = 0$ if  (\ref{gravitino_eom}) holds. 

It is actually possible to write down a simple expression for $T_m{}^{\mu}$, namely
\begin{equation}
T_m{}^{\mu} = \theta_m{}^{\mu} - \frac{e}{\kappa^2} (G_m{}^{\mu}(e,\omega(e,\psi)) - G_m{}^{\mu}(e,\omega(e)))
\end{equation}
The difference of the two Einstein tensors contains the total derivative terms (from $D_{\mu} \tau_{\nu} - D_{\nu} \tau_{\mu}$) and the torsion terms (from $\tau_{\mu} \tau_{\nu} - \tau_{\nu} \tau_{\mu}$) which we mentioned earlier. Because all terms in the action with gravitinos are Einstein and locally Lorentz invariant, $T_{\mu\nu}$ is symmetric on the spin 3/2 mass shell. On-shell, using (\ref{spin2_eom}), this reduces to $T_m{}^{\mu} = \frac{e}{\kappa^2} G_m{}^{\mu}(e,\omega(e))$, which is, of course, the usual Einstein equation.
Clearly, (\ref{eq:26}) is satisfied. (I thank M. Ro\v cek for a discussion.)

\section{Boundary terms in supergravity}

We now return to the issue of boundary terms. The following is work by two of
us (D.V. and P.v.N.). In addition to (\ref{bound_term}), there is a boundary term coming from the variation of $\omega$ in ${\cal L}_{EH}$
\begin{equation}
\partial_{\mu} \left[-\frac{e}{\kappa^2}\, e \, e_m{}^{\mu} e_n{}^{\nu} \delta_{\rm susy} \omega_{\nu}{}^{mn} \right] \label{omega_bound_term}
\end{equation}
For the field equations there are also boundary terms from varying $\omega$ and from varying $\psi$
\begin{equation}
\partial_{\mu} \left[-\frac{e}{2\kappa^2}\, e\, e_m{}^{\mu} e_n{}^{\nu} \delta \omega_{\nu}{}^{mn} \right] \hspace{1cm} \mbox{and} \hspace{1cm} \partial_{\rho} \left[ -\frac{e}{2} \bar{\psi}_{\mu} \gamma^{\mu\rho\sigma} \delta \psi_{\sigma} \right] \label{other_bound_terms}
\end{equation}
In 1.5 order formalism, $\delta \omega_{\nu}{}^{mn}$ contains terms with $\delta e_{\rho}{}^{\mu}$ and terms with $\delta \psi_{\rho}$ by the chain rule. The question in now how to cancel the sum of (\ref{bound_term}) and (\ref{omega_bound_term}), as well as the terms in (\ref{other_bound_terms}) by BC and boundary actions which are consistent. This problem was studied in detail in \cite{Nieuwenhuizen:Vassilevich:2005} and we refer the reader to this article. Here we only give the main points.

For fermions one introduces projection operators
\begin{equation}
P_{\pm} = \frac{1}{2} (1 \pm \gamma^n) \hspace{40pt} \gamma^n = \gamma_m e_{\mu}{}^m n^{\mu} 
\end{equation} 
where $n^{\mu}$ is the normal to the boundary. It satisfies $\psi^{\dagger} (\gamma^n)^{\dagger} i \gamma^0 = -\bar{\psi} \gamma^n$  whether $n^{\mu}$ is spacelike, timelike or null. The local parameter is then restricted by $P_- \epsilon = 0$ (or $P_+ \epsilon = 0$), so half of local susy is lost at the boundary.

Let us first consider the spin 0, 1/2 system. For the spin 0 fields $S$ one finds the boundary variation $-S\, \partial_n S = 0$, so either the Dirichlet condition $S |_{\partial {\cal M}} = 0$ or the Neumann condition $\partial_n S |_{\partial {\cal M}} = 0$ must hold. Under susy $S$ varies into a spinor $\lambda$, and for $\lambda$ the Euler-Lagrange boundary term is $-\frac{1}{2} \lambda \gamma^n \delta \lambda$, where $\gamma^n = \gamma_{\mu} n^{\mu}$. The susy variations involve also a pseudoscalar $P$, and if $P_-\epsilon = 0$ the two sets of BC begin with
\begin{equation}
S |_{\partial {\cal M}} =0,\qquad \partial^n P|_{\partial {\cal M}} =0, \qquad P_-\lambda|_{\partial {\cal M}} =0
\label{set1-1}
\end{equation}
or 
\begin{equation} 
P |_{\partial {\cal M}} =0,\qquad \partial^n S |_{\partial {\cal M}} =0, \qquad P_+\lambda|_{\partial {\cal M}} =0 
\end{equation}
Susy variations of these conditions produce again BC with extra 
$\partial_n$ derivatives, but now these BC are conditions for off-shell
fields in the action. For example, consistency of the BC in (\ref{set1-1})
leads to the further set $\partial_n^{2m} S|_{\partial {\cal M}} =0$,
$\partial_n^{2m-1}P|_{\partial {\cal M}} =0$, $P_- \partial_n^{2m} \lambda |_{\partial {\cal M}} =0$
and $P_+ \partial_n^{2m+1}\lambda |_{\partial {\cal M}}=0$ for $m=1,2,3\dots$

One can then study all free spin 0, 1/2, 1, 3/2 and 2 systems which are rigidly susy. Again one finds two classes of consistent BC. For example, the first class contains $P_+ \psi_j = 0$ and Dirichlet conditions for  $A_n$ and Neumann conditions for $A_j$, and the other class has $P_- \psi_j = 0$ and the opposite for $A_{\mu}$. These BC are then generalized to the case of local susy. Einstein invariance implies (as always) that the diffeomorphism parameters $\zeta^{\mu}$ satisfies $\zeta^n |_{\partial {\cal M}} = 0$. The Gibbons-Hawking condition $\delta g_{ij} = 0$ implies for Einstein transformations that $\zeta_i$ is a Killing vector in the boundary, which in general implies also $\zeta_i |_{\partial {\cal M}} = 0$. The boundary condition on the gravitino which belongs to the same set as $\delta g_{ij} =0$ is $P_- \psi_n = 0$. A susy variation requires then for consistency that also $P_- \partial_n \epsilon = 0$. But then one has {\it both} Dirichlet and Neumann BC for $P_- \epsilon$. This excludes local susy transformations on the boundary. (In a BRST approach the local parameters become ghosts, and for them these BC are too strong.) The other solution, $K^{ij} |_{\partial{\cal M}} = 0$, seems consistent.

\section{Boundary terms in the quantum mass of solitons}

We now present some recent results on monopoles. First we quote from \cite{Rebhan:Nieuwenhuizen:Wimmer:2006} a summary of our previous work on solitons.

" The existence of
supersymmetric (susy) monopoles \cite{D'Adda:1978ur,D'Adda:1978mu} 
which saturate the Bogomolnyi bound \cite{Bogomolny:1976de}
also at the quantum level \cite{Witten:1978mh}
plays an important role in the successes of nonperturbative studies
of super Yang-Mills theories through dualities \cite{Seiberg:1994rs,Seiberg:1994aj,Alvarez-Gaume:1997mv,Tong:2005un}. 

On the other hand, a direct calculation of quantum corrections
to the mass and central charge of susy solitons has proved to
be fraught with difficulties and surprises.
While in the earliest literature it was assumed that supersymmetry
would lead to a complete cancellation of quantum corrections
to both \cite{D'Adda:1978ur,D'Adda:1978mu,Rouhani:1981id}, 
it was quickly realized
that the bosonic and fermionic quantum fluctuations
do not only not cancel, but have to match the infinities
in standard coupling and field renormalizations \cite{Schonfeld:1979hg,Kaul:1983yt,Chatterjee:1984xh,Yamagishi:1984zv,Uchiyama:1984kb,Imbimbo:1984nq}.
However, even in the simplest case of the 1+1 dimensional
minimally supersymmetric kink, there was until the end of the 1990's
an unresolved discrepancy in the literature as to the precise
value of one-loop contributions once the renormalization scheme has
duly been fixed. As pointed out in \cite{Rebhan:1997iv}, 
most workers had used regularization
methods which, when used naively, give inconsistent results
already for the exactly solvable sine-Gordon kink.
In the susy case, there is moreover an extra complication
in that the traditionally employed periodic boundary conditions
lead to a contamination of the results by energy located at
the boundary of the quantization volume, and the issue
of the correct quantum mass of the susy kink was finally
settled in Ref.~\cite{Nastase:1998sy} by the use of topological boundary
conditions, which avoid this contamination.\footnote{Ref.~\cite{Nastase:1998sy}
used ``derivative regularization'' to make this work. In 
mode regularization it turns out that one has to average over
sets of boundary conditions to cancel both localized
boundary energy and delocalized momentum 
\cite{Goldhaber:2000ab,Goldhaber:2002mx}.}
This singled out as correct the earlier result
of Ref.~\cite{Schonfeld:1979hg,Casahorran:1989vd}
and refuted the null results of 
Refs.~\cite{Kaul:1983yt,Chatterjee:1984xh,Yamagishi:1984zv,Uchiyama:1984kb}.
However it led to a new problem because it seemed 
that the central charge did not appear to receive corresponding quantum
corrections \cite{Imbimbo:1984nq}, which would
imply a violation of the Bogomolnyi bound.
In Ref.~\cite{Nastase:1998sy}
it was conjectured that a new kind
of anomaly was responsible, and in Ref.~\cite{Shifman:1998zy}
Shifman, Vainstein, and Voloshin 
subsequently demonstrated that supersymmetry requires an anomalous
contribution to the central charge current.\footnote{Refs.\
\cite{Graham:1998qq,Casahorran:1989nr}, who had obtained
the correct value for the quantum mass also claimed
a nontrivial quantum correction
to the central charge apparently without the need of the anomalous term
proposed in Ref.~\cite{Shifman:1998zy}. However,
as shown in Ref.~\cite{Rebhan:2002yw}, this was achieved by formal
arguments handling ill-defined since unregularized quantities.}
The latter appears
in the same multiplet as the trace and conformal-susy anomalies,
and ensures BPS saturation even in the $N=1$ susy kink,
where initially standard multiplet shortening arguments seemed not
to be applicable.\footnote{That multiplet shortening also
occurs in the $N=1$ susy kink was eventually clarified
in Ref.\ \cite{Losev:2000mm}.}

In Ref.~\cite{Rebhan:2002uk} we have developed a version
of dimensional regularization which can be used
for solitons (and instantons).
The soliton is embedded in a higher-dimensional space
by adding extra trivial dimensions \cite{Luscher:1982wf,Parnachev:2000fz}
and choosing a model which is supersymmetric in the bigger space
and which reproduces the original model by dimensional reduction.
This is thus a combination of standard 't Hooft-Veltman
dimensional regularization \cite{'tHooft:1972fi} which goes up in dimensions,
and susy-preserving dimensional reduction
\cite{Siegel:1979wq,Capper:1980ns}, which goes down.
In Ref.~\cite{Rebhan:2002yw}
we demonstrated how the {anomalous contribution to} the central charge
of the susy kink can be obtained as a remnant of parity violation
in the odd-dimensional model used for embedding the susy kink,
and recently we showed that the same kind of anomalous contribution
arises in the more prominent case of the 3+1-dimensional monopole
of $N=2$ super-Yang-Mills theory in the Higgs\footnote{Anomalous
contributions to the central charge appear also in the newly
discovered ``confined monopoles'' pertaining to the
Coulomb phase \cite{Auzzi:2003fs,Tong:2003pz,Shifman:2003uh,Shifman:2004dr,Hanany:2004ea,Eto:2006pg}, 
which turn out to be related to central charge
anomalies of 1+1-dimensional $N=2$ sigma models with twisted mass
\cite{Shifman:2006bs}.}
phase \cite{Rebhan:2004vn}.
This previously overlooked \cite{Kaul:1984bp,Imbimbo:1985mt}
finite contribution turns out to be in fact essential 
for consistency of these direct calculations
with the $N=2$ low-energy effective action
of Seiberg and Witten 
\cite{Seiberg:1994rs,Seiberg:1994aj,Alvarez-Gaume:1997mv}.
(We
have also found previously overlooked finite contributions to
both mass and central charge of the $N=2$ vortex in 2+1 dimensions
\cite{Vassilevich:2003xk,Rebhan:2003bu},
which are however not associated with conformal anomalies
but are rather standard renormalization effects.) "

We now discuss the boundary terms in the 1-loop corrections to the mass of a monopole in $N=2$ \cite{Rebhan:Nieuwenhuizen:Wimmer:2004} and $N=4$ \cite{Rebhan:Nieuwenhuizen:Wimmer:2006} super Yang-Mills. We define the stress tensor by $T_{\mu\nu} = - 2 \frac{\delta}{\delta g^{\mu\nu}} S$.  The basic idea is that the gravitational stress tensor for bosons is of the form $\partial \phi \, \partial \phi$, but the sum over zero point energies comes from terms of the form $- \phi \, \partial \partial \phi$. Clearly, the mass consists of a sum over zero point energies (which is what one usually assumes) {\it and} surface terms $\phi\, \partial_n \phi$, and the latter contribute if there are massless fields in the theory. There are no surface terms for fermions. 

For $N=4$ super Yang-Mills theory with a gravitationally background-covariant gauge-fixing term
\begin{equation}
{\cal L}_{\rm fix} = - \frac{1}{2} \frac{1}{\sqrt{-g}} (D_{\mu}(A) \sqrt{-g} g^{\mu\nu} a_{\nu})^2
\end{equation}
where $A_{\mu}$ is a background field and $a_{\mu}$ the quantum field, the one-loop surface terms in $T_{00}$ which are quadratic in quantum fields read
\begin{eqnarray}
M^{(1)\rm surf} &\hspace{-5pt} = \hspace{-5pt}& \int d^3x\, T_{00}^{(2){\rm tot. deriv.}} \nonumber \\
&\hspace{-5pt} = \hspace{-5pt}& \int d^3x \left[ \frac{1}{4} \partial_j^2 \langle a_0^2+a_{\bf S}^2 +2bc \rangle -\frac{1}{2} \partial_j\partial_k \langle a_j a_k\rangle +2\partial_j \langle a_j \partial_0 a_0 \rangle \right]
\end{eqnarray}
where $a_j$, $a_0$, $a_{\mu}$ are spin 1 fields and $a_5, ... \, , a_{10}$ spin 0 fields with $j,k=1,2,3$ and $b$, $c$ are the antighosts and ghosts. The difference between $N=2$ and $N=4$ is
only the range of the index ${\bf S}\,$: ${\bf S}=1,2,3,5,6$ for
$N=2$, and ${\bf S}=1,2,3,5,\ldots,10$ for $N=4$.
We shall show that the sum of all surface contributions
cancels for the $N=2$ monopole, but for $N=4$ the extra four 
(pseudo-)scalars
yield a new type of divergence which we will have to dispose of.

The exact propagators in a monopole background are very complicated, but fortunately we only need their form at $r \rightarrow \infty$. Then one finds that for the $N=2$ model the surface terms cancel 
\begin{eqnarray}
M^{(1)\rm surf,N=2} &\hspace{-5pt} = \hspace{-5pt}& \lim_{r\to\infty}\frac{1}{4} 4\pi r^2
\frac{\partial}{\partial r} \langle a_0^2+a_j^2+s^2+p^2+2bc-2s^2 \rangle \nonumber \\
&\hspace{-5pt} = \hspace{-5pt}& (-1+3+1+1-2-2)\lim_{r\to\infty}\pi r^2 \frac{\partial}{\partial r} \langle s^2 \rangle = 0
\end{eqnarray}
where $s$ and $p$ are the two scalar fields of pure $N=2$ susy Yang-Mills theory.
On the other hand, in the $N=4$ case there are 4 extra spin zero fields, and their contribution to the surface terms is not only nonvanishing, but even divergent (!)
\begin{equation}
M^{(1)\rm surf,N=4}=4\times \frac{1}{4} \lim_{r\to\infty}4\pi r^2 \frac{\partial}{\partial r}\langle s^2\rangle
\end{equation}
\begin{equation}
\langle s^2\rangle =G^{aa}(y,x)|_{y=x}
\simeq 2\,\langle y|\frac{-i}{-\Box+m^2-\frac{2m}{r}}|x\rangle |_{y=x}+{\rm const.}
\end{equation}
The factor 2 comes from a trace over $\delta^{ab} - \hat{x}^a \hat{x}^b$. Inserting momentum eigenstates yields
\begin{eqnarray}
\langle s^2\rangle &\simeq& 2\int \frac{d^{4+\epsilon} k}{(2\pi)^{4+\epsilon}} \frac{-i}{(k^2+m^2)+2ik^\mu\partial_\mu-\partial_\mu^2 -\frac{2m}{r}} \nonumber \\
&\simeq & 2\, \frac{2m}{r} \int \frac{d^{4+\epsilon} k}{(2\pi)^{4+\epsilon}}\frac{-i}{(k^2+m^2)^2}
=4\frac{m}{r}I
\end{eqnarray}

Fortunately, there is another divergence not yet used: an overall $Z$ factor for the stress tensor \cite{Rebhan:2005yi} (in addition to the usual wave function and coupling renormalization factors, which appear because we use a non-supersymmetric gauge-fixing term). It now turns out that the $Z$ factor for the renormalization of the composite operators (the currents) {\it cancels} the surface corrections in the $N=4$ model.

This $Z$ factor was obtained by considering a simple matrix element in the spontaneously broken but topologically trivial sector of the (ordinary, non-conformal) current for the magnetic monopole at vanishing incoming momentum, and constructing a counterterm which subtracts both the divergent and finite contributions. This matrix element had as external fields a massless vector and scalar background field, and the counterterm turned out to be part of a conformal multiplet of improvement currents. So the non-conformal currents can be split into conformal currents minus improvement terms, and only the latter renormalize, and they renormalize multiplicatively. A well-known theorem says that conserved currents do not renormalize (by an overall $Z$ factor), but this only applies to internal symmetries and not to stress tensors and other currents of spacetime symmetries.

A subtle point, however, remains. One could have started with the improved stress tensor \cite{Callan:1970ze}. This time there is no overall $Z$ factor for the currents. The classical value of the monopole mass is then 2/3 of the previous (conventional, unimproved) value. Which stress tensor should one use to obtain the mass of a monopole? The $N=4$ model is conformal, and in the AdS-CFT program one uses conformal currents. On the other hand, dimensional reduction from 10 to 4 dimensions yields non-conformal currents. In flat space both stress tensors, and in fact both current multiplets, are consistent. Can one decide whether to use the ordinary (unimproved) or the improved (conformal) stress tensor to obtain the monopole mass by consistency arguments involving gravity? Hopefully, by the next Puri meeting we will know.

\vspace{1cm}

{\bf Acknowledgements:} I thank G. Dat\'{e} for inviting me for a stay at Chennai (Madras) before this conference, and R. Kaul, D. Indumathi, M.V.N. Murthy and G. Rajasekaran for very useful discussions at Chennai and Puri.

\providecommand{\href}[2]{#2}\begingroup\raggedright

 \end{document}